# FPGA Implementation of Robust Residual Generator

Young-Man Kim

Liberty University, Lynchburg, USA
(Email: ykim81@liberty.edu)

*Abstract*— **In this paper, one can explicitly see the process of implementing the robust residual generator on digital domain, especially on FPGA. Firstly, the baseline model is developed in double precision floating point format. To develop the baseline model, key parameters such as SNR and detection window length are selected in the identification stage. SNR is far more important in determining the residual generator's performance in the sense of FAR even though the detection window length can reduce the bias effect. From that simulation, the proper value of input signal and detection window size can be determined. Then, one can find the proper format of fixed-point data type by simulating it. In this research, Matlab HDL coder is used to generate HDL codes and the generation report proposes fixed point data format. The proper value is selected to satisfy desired FAR and verified. The digitized implementation of robust residual generator on FPGA opens the doors to better fault detection at fast speed.**

I. Introduction

Fault detection is an important topic in control community as systems become more complex and early detection is critical to prevent any disaster. For effective fault detection, a signal which alarms one without false should be generated. That signal is commonly called residual, and it is typically generated by finding the difference between the output of the actual system and that of the reference model [1-3,5]. In some cases, the reference model can be developed based on scientific theory but, if it is not the case, the reference model is identified using various techniques. Among them, the PBSID (Predictor-Based System Identification) technique [6] has gained its popularity because it can be easily connected to fault detection, fault tolerant control [11, 16, 17], online monitoring, predictive maintenance and all those disciplines can be categorized as data-driven modeling [7].

Recently, more interests have converged on online fault detection due to the necessity of real time monitoring with the advancement of digital technology. For online fault detection, the residual signal needs to be recursively updated [4] and it requires fast computation with minimal errors. For online recursive update of residual, fast handling of many I/O data is necessary. For fast handling, the setup of fault detection algorithm can be effectively implemented by using Markov parameters instead of using model parameters such as (A, B, C, D) matrices in state space domain [8]. Because Markov parameters are composed of collected I/O data, it enables the handling process better suited to online monitoring. However, initial identification of Markov parameter requires many I/O data collection, and the size of data needs to be large enough to avoid the bias-effect of the finite data size [9]. In addition, the collected I/O data is contaminated with noise from measurement and process [6]. Unbiased residual generation which is robust to measurement and process noise is discussed at [10]. Reference [10] proposes the tuning method to find a proper detection window size to minimize the bias effect. It also compensates for the influence of noise on the residual generation to minimize false alarm [12] and maximize the robustness to the noise. In this research the relationship between the SNR (Signal to Noise Ratio) value and the efficacy of fault detection algorithm is searched using simulation.

To detect any changes in residual, proper test statistic and threshold should be selected. In this research, Chi-squared ($\chi^2$) one is used, and it is one of the whitening-based change detection methods. In [10], the non-centrality of Chi-squared distribution relates to the induced bias due to the identification error. In this research, the relationship between the non-centrality due to the bias from finite detection window and the efficacy of fault detection algorithm is searched using simulation.

For fast computation, residual generation using Markov parameter is useful as mentioned above. A digital computer or an embedded system is used for the implementation of fault detection algorithms. If one implements specific parts of the algorithm on FPGA (Field Programmable Gate Array), one can expect more speed enhancement through hardware acceleration. The implementation on digital domain (for example, embedded systems, SoC (System on Chip), FPGA, etc.) benefits from using fixed-point data type. Fixed point data type is useful for fast processing even though it has limited accuracy compared with floating point [13]. In addition, the format of fixed-point, $Q_{m.f}$ ($m$: word length, $f$: fractional length) affects the efficacy. With Matlab, the baseline model is developed in double precision and used to compare with the implementation in fixed point data type. In this research, the relationship between the fixed-point format and the efficacy of fault detection algorithm is searched using simulation.

For test statistic, Chi-squared one, $\chi^2$, is used. It provides values which can be used as a threshold and the right tail probability is used as false alarm rate, $\alpha$.

In this research, with Matlab, a baseline model of robust residual generator is developed to check the impact of SNR and detection window on $\alpha$ based on $\chi^2$. Then, it is implemented with fixed point data type to check its efficacy on digital domain. The $m$ and $f$ value of $Q_{m.f}$ is sought to satisfy $\alpha$.

The rest of the paper is organized as follows. In section II, the robust residual generator in state space form is set up for laying the theoretical background [10]. It is formulated in VARX (Vector Auto Regressive with eXogenous input) form to make the parameter setting easy with Markov parameter. Then, a robust residual generator is constructed from considering a system identification error and it is recursively updated for online fault detection. Section II is a condensed version of the result in [10] and one can refer to it for more details. In section III, using Matlab, a baseline model is developed for checking the impact of SNR and detection window on false alarm rate, $\alpha$. Overall result shows that the inclusion of identification error into residual generation with the proper consideration of SNR and detection widow during the identification process enhances the quality of fault detection algorithm. In section IV, the fixed-point implementation of $\chi^2$ is described. The selection of data format such as word length and fractional length and its impact on detection performance is discussed using a simulation result. In section V, the algorithm is implemented on FPGA board to check the possibility of fast real-time computation with some hardware acceleration. Then, the conclusion follows.

The notations used in this paper are standard. $R^{m \times n}$ represents a real matrix with $(m, n)$ size. $\chi^2$ represents a Chi-squared distribution. $M^\dagger$ represents the pseudo-inverse of matrix $M$. The operator $\otimes$ stands for the Kronecker product. $\hat{x}$ represents the estimates of the state vector, $x$.

## II. SETUP OF ROBUST RESIDUAL GENERATOR

### A. VARX Model for Ouput Estimator

One considers the following model for fault detection:

$$x(k+1) = Ax(k) + Bu(k) + Ef(k) + Fw(k) \quad (1)$$

$$y(k) = Cx(k) + Du(k) + Gf(k) + v(k) \quad (2)$$

where $x(k) \in R^n$, $y(k) \in R^l$, $u(k) \in R^m$, $f(k) \in R^{n_f}$, $w(k) \in R^{n_w}$, $v(k) \in R^l$. $f$ represents fault signals, $w$ is the process noise, and $v$ is the measurement noise. $w$ and $v$ are assumed to be white zero-mean Gaussian. One assumes that the system (1,2) is internally stable with or without closed-loop stabilizing controller.

Using innovation signal, $e(k) = y(k) - C\hat{x}(k) - Du(k)$, (1,2) can be reformulated as follows.

$$\hat{x}(k+1) = A\hat{x}(k) + Bu(k) + Ke(k), \quad (5)$$

$$y(k) = C\hat{x}(k) + Du(k) + e(k). \quad (6)$$

where $K$ is Kalman gain and $e$ is white Gaussian with covariance matrix, $\Sigma_e$. From (5,6), a closed loop system equation can be derived as follows.

$$\hat{x}(k+1) = (A - KC)\hat{x}(k) + (B - KD)u(k) + Ky(k), \quad (7)$$

$$y(k) = C\hat{x}(k) + Du(k) + e(k). \quad (8)$$

where $\Phi \triangleq A - KC$, $\tilde{B} = B - KD$. One assumes that $\Phi$ is stable.

Using (7,8), one can derive a VARX model to set up the output equation with Markov parameters from starting $k - p$ time instant by recursively solving for $p$ sampling times as shown below.

$$\hat{x}(k) = \Phi^p \hat{x}(k-p) + \sum_{\tau=0}^{p-1} \Phi^\tau [\tilde{B} \quad K] \begin{bmatrix} u(k-\tau-1) \\ y(k-\tau-1) \end{bmatrix}, \quad (9)$$

$$y(k) = C\Phi^p \hat{x}(k-p) + \sum_{\tau=0}^{p-1} C\Phi^\tau [\tilde{B} \quad K] \begin{bmatrix} u(k-\tau-1) \\ y(k-\tau-1) \end{bmatrix}$$

$$+ [D \quad 0] \begin{bmatrix} u(k) \\ y(k) \end{bmatrix} + e(k). \quad (10)$$

As (9,10) shows, the system output can be determined from $\{D, C\Phi^j \tilde{B}, C\Phi^j K, j = 0, \ldots, p-1\}$ which are called the Markov parameters. To identify Markov parameters, one collects $N$ ($p \ll N$) data into a block row vector denoted as $Y_{id}$ as below.

$$\underbrace{[y(k) \; y(k+1) \cdots y(k+N-1)]}_{Y_{id}}$$

$$= C\Phi^p \cdot \underbrace{[\hat{x}(k-p) \; \hat{x}(k-p+1) \cdots \hat{x}(k+N-p-1)]}_{X_{id}}$$

$$+ \underbrace{[C\Phi^{p-1}\tilde{B} \; C\Phi^{p-1}K \cdots C\tilde{B} \; CK \mid D]}_{\Xi}$$

$$\cdot \underbrace{\begin{bmatrix} u(k-p) & u(k-p+1) & & u(k+N-p-1) \\ y(k-p) & y(k-p+1) & \cdots & y(k+N-p-1) \\ \vdots & \vdots & \vdots & \vdots \\ u(k-1) & u(k) & \cdots & u(k+N-2) \\ y(k-1) & y(k) & \cdots & y(k+N-2) \\ u(k) & u(k+1) & & u(k+N-1) \end{bmatrix}}_{Z_{id}}$$

$$+ \underbrace{[e(k) \; e(k+1) \cdots e(k+N-1)]}_{E_{id}}. \quad (11)$$

Equation (11) can be simply written as:

$$Y_{id} = C\Phi^p X_{id} + \Xi Z_{id} + E_{id}. \quad (12)$$

where $\Xi$ is the Markov parameter. As one assumes the stability of (12) with $p \to \infty$, the estimate of Markov parameter, $\hat{\Xi}$, can be found below with the assumption that $Z_{id}^\dagger$ has full row rank.

$$\hat{\Xi} = Y_{id} \cdot Z_{id}^\dagger. \quad (13)$$

Thus, the Markov parameter estimate error is found as below.

$$\Delta\hat{\Xi} \triangleq \Xi - \hat{\Xi} = C\Phi^p X_{id} Z_{id}^\dagger + E_{id} Z_{id}^\dagger. \quad (14)$$

With $p \to \infty$, $\Delta\hat{\Xi}$ can be simplified as below.

$$\Delta\hat{\Xi} = E_{id} Z_{id}^\dagger. \quad (15)$$

After identifying the Markov parameter as (13), one collects $L$ I/O data and innovations to lump them into a column vector to form a VARX as follows.

$$\underbrace{\begin{bmatrix} y(k-L+1) \\ y(k-L+2) \\ \vdots \\ y(k) \end{bmatrix}}_{y_{k,L}} = \underbrace{\begin{bmatrix} C\Phi^p \hat{x}(k-L-p+1) \\ C\Phi^p \hat{x}(k-L-p+2) \\ \vdots \\ C\Phi^p \hat{x}(k-p) \end{bmatrix}}_{b_{k,L}}$$

$$+ \underbrace{\begin{bmatrix} C\Phi^{p-1}\tilde{B} & C\Phi^{p-1}K & \cdots & \cdots & C\tilde{B} & CK \\ 0 & 0 & C\Phi^{p-1}\tilde{B} & C\Phi^{p-1}K & \cdots & C\Phi\tilde{B} & C\Phi K \\ \vdots & \vdots & & & \vdots & \vdots \\ 0 & 0 & 0 & C\Phi^{p-1}\tilde{B} & \cdots & \cdots & C\Phi^{L-1}K \end{bmatrix}}_{H_z^{L,p}}$$

$$\cdot \underbrace{\begin{bmatrix} u(k-L-p+1) \\ y(k-L-p+1) \\ \vdots \\ u(k-L) \\ y(k-L) \end{bmatrix}}_{z_{k-L,p}}$$

$$+\begin{bmatrix} D & 0 & \cdots & 0 \\ C\tilde{B} & D & \cdots & 0 \\ \vdots & \vdots & \vdots & 0 \\ C\Phi^{L-2}\tilde{B} & C\Phi^{L-3}\tilde{B} & \cdots & D \end{bmatrix}_{T_u^L} \underbrace{\begin{bmatrix} u(k-L+1) \\ \vdots \\ u(k) \end{bmatrix}}_{u_{k,L}}$$

$$+\begin{bmatrix} 0 & 0 & \cdots & 0 \\ CK & 0 & \cdots & 0 \\ \vdots & \vdots & \vdots & 0 \\ C\Phi^{L-2}K & C\Phi^{L-3}K & \cdots & 0 \end{bmatrix}_{T_y^L} \underbrace{\begin{bmatrix} y(k-L+1) \\ \vdots \\ y(k) \end{bmatrix}}_{y_{k,L}}$$

$$+\underbrace{\begin{bmatrix} e(k-L+1) \\ e(k-L+2) \\ \vdots \\ e(k) \end{bmatrix}}_{e_{k,L}}$$

(16)

where $L$ is for the detection horizon, $p$ for the past one, and $z(k)$ is a new notation for lumped I/O as defined below.

$$z(k) = [u^T(k) \quad y^T(k)]^T. \qquad (17)$$

Thus, (16) can be written in lumped VARX form as below.

$$y_{k,L} = b_{k,L} + H_z^{L,p} z_{k-L,p} + T_u^L u_{k,L} + T_y^L y_{k,L} + e_{k,L}. \qquad (18)$$

As one can notice from (12,16), the matrices in (16) such as $H_z^{L,p}, T_u^L, T_y^L$ can be found using (13). The residual generator for fault detection can be defined as the difference between the measured output ($y_{k,L,meas}$) and the calculated one ($y_{k,L,calc}$) of (18) as follows.

$$r_{k,L} = y_{k,L,meas} - y_{k,L,calc}$$
$$= y_{k,L,meas} - T_y^L y_{k,L} - H_z^{L,p} z_{k-L,p} - T_u^L u_{k,L}. \qquad (19)$$

To apply $\chi^2$ test statistic, one needs to find the covariance of the residual signal (19). One can first investigate the I/O and the Markov parameter relationship for the output $y(k - L + 1)$. It can be found from (16) as follows.

$$y(k - L + 1) = C\Phi^p \hat{x}(k - L - p + 1)$$
$$+ \Xi \begin{bmatrix} z_{k-L,p} \\ u(k-L+1) \end{bmatrix} + e(k - L + 1). \qquad (20)$$

Using (14), one can replace the true Markov parameter $\Xi$ by its estimate $\hat{\Xi} + \Delta\hat{\Xi}$. Then, (21) below is the stochastic term in (20).

$$\Delta\hat{\Xi} \cdot \begin{bmatrix} z_{k-L,p} \\ u(k-L+1) \end{bmatrix} + e(k - L + 1). \qquad (21)$$

Thus, the covariance of $r_{k,L}$, cov ($r_{k,L}$), can be derived from (21) as follows.

*Theorem 1* (Ref. to [10]): The covariance of the residual $r_{k,L}$, $\Sigma_{\Delta\hat{\Xi},e}^L$, can be approximated by the following matrix:

$$\Sigma_{\Delta\hat{\Xi},e}^L = \Sigma_{\Delta\hat{\Xi}}^L + \Sigma_e^L = (Z_{ol}^T(Z_{id}Z_{id}^T)^{-1}Z_{ol} + I_L) \otimes \Sigma_e, \qquad (22)$$

where

$$Z_{ol} = \begin{bmatrix} z_{k-L,p} & z_{k-L+1,p} & \cdots & z_{k-L,p} \\ \hline u(k-L+1) & u(k-L+2) & \cdots & u(k) \end{bmatrix}$$

and $\otimes$ is the Kronecker product. (23)

To define the Chi-squared statistic, $\tau(k)$, one needs to whiten the residual as follows [14].

$$\tilde{r}_{k,L} \triangleq \left(\Sigma_{\Delta\hat{\Xi},e}^L\right)^{-\frac{1}{2}} r_{k,L}. \qquad (24)$$

Using (24), the test statistic can be defined as below.

$$\tau(k) \triangleq \|\tilde{r}_{k,L}\|_2^2 = \left\|\left(\Sigma_{\Delta\hat{\Xi},e}^L\right)^{-\frac{1}{2}} r_{k,L}\right\|_2^2. \qquad (25)$$

Thus, the fault detection test can be defined using the Chi-squared statistic as in (26) below where $\gamma_\alpha$ is the threshold selected from $\chi^2$ table depending on DOF, and it is $(L-1) \cdot l$ in this research. $\alpha$ is the false alarm rate (FAR) which is the right tail probability of $\chi^2$ pdf.

$$\tau(k) > \gamma_\alpha \text{ if a fault occurs.} \qquad (26)$$

The non-centrality of $\chi^2$ distribution is resulted from the bias term $b_{k,L}$ in (16) but, through this research, one notices that it is omitted from the assumption that $\Phi$ is stable and the past window for identification, $p$, and the detection window, $L$, can be big enough such that its impact on test statistic is minimal compared with that of the covariance (22).

### III. DEVELOPMENT OF BASELINE MODEL FOR A FIXED-POINT IMPLEMENTATION OF A ROBUST RESIDUAL GENERATOR

To implement robust residual generator on digital domain, especially FPGA (Field Programmable Gate Array), a baseline model is designed using double precision floating point format with Matlab. It can be used to check the performance of digitally implemented residual generator by comparison. Two parameters such as detection window size and SNR in identification stage play key roles in determining the performance of fault detection. The bigger the detection window size is, the smaller the bias is as described in section II and SNR affects the quality of Markov parameter identification as shown in (15,21). Using the following SISO system, one develops the baseline model of robust residual generator which is used as a reference for developing its fixed-point implementation on digital domain, especially FPGA.

$$y(k) = d \cdot u(k) + e(k) \qquad (27)$$

where $e(k)$ is a zero-mean white Gaussian noise with variance $\sigma_e^2$ and $d$ is a static gain. Firstly, $L$ I/O samples are collected to identify the unknown static gain $d$. The estimate, error, and variance are defined as follows.

$$\hat{d} = Y_{id} \cdot U_{id}^\dagger, \qquad (28)$$

$$\Delta\hat{d} = d - \hat{d} = E_{id} U_{id}^\dagger, \qquad (29)$$

$$\text{var}(\Delta\hat{d}) = \frac{\sigma_e^2}{U_{id} U_{id}^\dagger} \qquad (30)$$

where $E_{id}, U_{id} \in R^{1 \times N}$ and its variance has the same form of (13,15). SNR is defined as follows.

$$SNR = 10 \cdot \log_{10} \frac{\frac{1}{(L-1)} U_{id} U_{id}^\dagger}{\sigma_e^2}. \qquad (31)$$

The residual and its characteristics can be described as below.

$$r_{k,L} = y_{k,L,meas} - y_{k,L,calc}$$
$$= d \cdot u(k) + e(k) - \hat{d}u(k) = \Delta\hat{d} \cdot u(k) + e(k) \qquad (32)$$

$$\text{var}(r(k)) = \text{var}(\Delta\hat{d} \cdot u(k)) + \text{var}(e(k))$$

$$= u^2(k) \cdot \text{var}(\Delta \hat{d}) + \sigma_e^2$$

$$= \frac{u^2(k) \cdot \sigma_e^2}{U_{id} U_{id}^T} + \sigma_e^2. \quad (33)$$

To accommodate any faults, one can rewrite (32) as follows.

$$r_{k,L} = \Delta \hat{d} \cdot u(k) + e(k) + f(k). \quad (34)$$

Thus, $\frac{r_{k,L}^2}{var(r_{k,L})}$ is $\chi^2$ distributed with $(L-1)$ DOFs. Thus, if the following test static, $\tau(k)$, is bigger than the threshold which is selected with the desired FAR from $\chi^2$ distribution table, then fault occurs with $\alpha$ and it can be described as below.

$$\tau(k) = \frac{r_{k,L}^2}{var(r_{k,L})} > \gamma_\alpha. \quad (35)$$

As one can see from (35), the test statistic takes the online information into account. Simulation is conducted to check the impact of selection of two key parameters on fault detection performance. For simulation, these values are set for the parameters: $d = 2, \sigma_e = 1$. The detection window, $L$, and SNR are varied. Fig. 1 below shows the result of simulation. The performance is defined that $\alpha$ is less than 0.5%. Additive-type faults are injected between 400 and 700 sample time and the resultant $\tau(k)$ is plotted in Fig. 2. As one can see, the longer detection window does not improve the performance if SNR is below 0 [dB]. Moreover, if the SNR is too low, the increment of detection window does not improve the performance at all. Thus, the consideration of SNR during identification is more critical factor than the selection of detection window length for better performance. At Fig. 2, the threshold selection from reading it using the $\chi^2$ distribution table is 38.6. It shows that the double precision floating point implementation of the generator can do excellent job at robust fault detection.

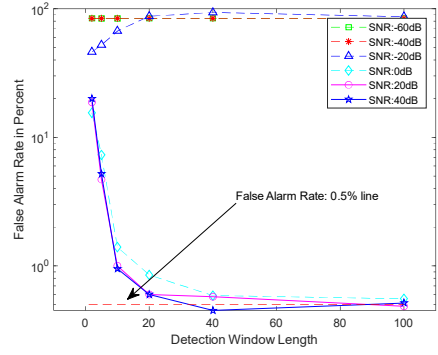

Fig. 1. False alarm rate w.r.t detection window length

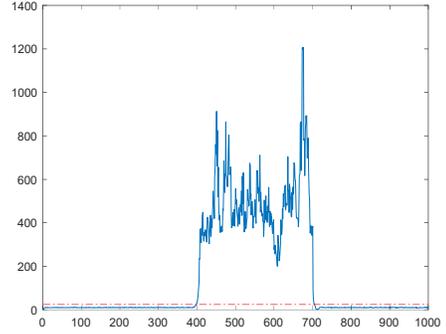

Fig. 2. Test statistic with threshold line

Table 1 shows how much SNR affects the estimation error.

Table 1. SNR w.r.t Estimation Error

| SNR [dB] | $\Delta \hat{d}$ |
|---|---|
| -60 | 141.89 |
| -40 | -7.69 |
| -20 | -0.65 |
| 0 | -0.10 |
| 20 | 0.01 |
| 40 | -0.002 |

As one can easily imagine, for faster detection, the detection window length should be as small as possible. However, it can cause an increase in FAR. Thus, from this simulation, detection window is twenty with SNR 20 [dB] can make the most desirable FAR result.

### IV. FIXED POINT IMPLEMENTATION OF ROBUST RESIDUAL GENERATOR

After finding a baseline model, fixed-point implementation can be conducted. Compared with floating point, fixed-point implementation is more favorable on digital domain due to its fast processing. In this research, the digital domain is specifically meant to be FPGA, but it can be applicable to any other means.

Fixed point data format, $Q_{m,f}$ ($m$: word length, $f$: fractional length), is checked how it can impact on detection performance. In other words, one seeks proper value of $m$ and

$f$ to satisfy the desired FAR. The Matlab HDL coder is used for further investigation into the choice of $m$ and $f$. The code generation is implemented on the selected FPGA board, Xilinx PYNQ-Z2 [15]. One can generate Hardware Description Language (HDL) codes such as VHDL, Verilog, etc., of the baseline model with user-defined testbench code. The test report suggests their proposed values which satisfy the desired FAR. Table 2. below shows the proposed values for each variable. The common fractional length $f$ for used variables which satisfy the selected FAR of 0.5% is six.

Table 2. List of Variables and Their Proposed Type of Fixed Data Format to Satisfy the Selected FAR

| Variable Name | Type | Sim Min | Sim Max | Static Min | Static Max | Whole Number | ProposedType (Best For FL = 6) |
|---|---|---|---|---|---|---|---|
| Chi_sq | double | 0.7077550627653921 | 1557.684152595177 | | | No | numerictype(0, 17, 6) |
| N | double | 10 | 10 | | | Yes | numerictype(0, 4, 0) |
| count | double | 1 | 2001 | | | Yes | numerictype(0, 11, 0) |
| dhat | double | 2.04 | 2.04 | | | No | numerictype(0, 8, 6) |
| i | double | 1 | 10 | | | Yes | numerictype(0, 4, 0) |
| r | double 10 x 1 | -8.063083634712196 | 16.232548989745425 | | | No | numerictype(1, 12, 6) |
| r_avg | double | -3.652777233214148 | 10.692100391687019 | | | No | numerictype(1, 11, 6) |
| r_sq | double 10 x 1 | 0 | 263.4956467044852 | | | No | numerictype(0, 17, 6) |
| r_sq_sum | double | 0 | 1172.7825383849695 | | | No | numerictype(0, 17, 6) |
| r_sub_ravg | double 10 x 1 | -11.288388915769156 | 13.783910814366264 | | | No | numerictype(1, 11, 6) |
| r_sub_ravg_sq | double 10 x 1 | 0 | 189.99610733840325 | | | No | numerictype(0, 14, 6) |
| r_sub_ravg_sq_sum | double | 0 | 454.8409417261664 | | | No | numerictype(0, 15, 6) |
| r_sum | double | -36.52777233214148 | 106.92100391687019 | | | No | numerictype(1, 14, 6) |
| r_var | double | 0 | 45.48409417261664 | | | No | numerictype(0, 12, 6) |
| u | double | 2 | 2 | | | Yes | numerictype(0, 2, 0) |
| ym | double | -3.983083634712197 | 20.312548989745427 | | | No | numerictype(1, 12, 6) |

Table 3. shows the summary of the implemented digital elements on PYNQ board.

Table 3. Summary of the Implemented Digital Elements on PYNQ-Z2 Board

| Summary | |
|---|---|
| Multipliers | 4 |
| Adders/Subtractors | 26 |
| Registers | 44 |
| Total Register Bits | 577 |
| RAMs | 0 |
| Multiplexers | 5 |
| I/O Bits | 43 |
| Shifters | 0 |

Fig. 3 shows the resultant $\tau(k)$ with threshold. It is implemented with $f = 6$ which is proposed by HDL coder, and it satisfies $\alpha < 0.5\%$ with some false alarms.

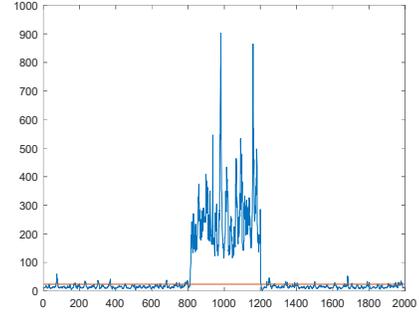

Fig. 3. Fault detection with FAR less than 0.5% by implementing the robust residual generator on PYNQ-Z2 FPGA board.

## V. FPGA IMPEMENTATION

Using Vitis HDL and PYNQ tools, the developed algorithm is implemented on Xilinx Zynq Z2 board both in floating point and fixed point. The floating point is set as single data type and fixed point is set based on the findings in section IV. As one can easily notice, the mostly used computation in residual calculation is the matrix-vector multiplication with addition. The programmable fabric inside the board has several DSP blocks which are used for floating point implementation. For speedy calculation with FPGA fabric, a pipeline is inserted. A memory block called BRAM (Block RAM) in the board is used for saving a matrix data. Without considering the overhead time for accessing BRAM memory, the floating-point implementation costs 426 [cycles] and the fixed point one costs 36 [cycles]. The unit latency time is 7.3 [ns] for 1 [cycle] with the maximum achievable frequency of 136 [MHz]. The speed improvement in fixed-point implementation is already expected. The comparison of used resources is in the following table.

Table 4. Resource Usage

| | FF | LUT | DSP |
|---|---|---|---|
| Fixed point | 1056 | 2036 | 0 |
| Floating point | 2032 | 3252 | 1 |

## VI. CONCLUSION

In this paper, one can explicitly see the process of implementing the robust residual generator on digital domain, especially on FPGA. Firstly, the baseline model is developed in double precision floating point format. To develop the baseline model, key parameters such as SNR and detection window length are selected in the identification stage. SNR is far more important in determining the residual generator's performance in the sense of FAR even though the detection window length can reduce the bias effect. From that simulation, the proper value of input signal and detection window size can be determined. Then, one can find the proper format of fixed-point data type by simulating it. In this research, Matlab HDL coder is used to generate HDL codes and the generation report proposes fixed point data format. The proper value is selected to satisfy desired FAR and verified. The digitized implementation of robust residual generator on FPGA opens the doors to better fault detection at fast speed.